\documentclass[aps,pra,twocolumn,showpacs,reprint,groupedaddress,superscriptaddress,longbibliography]{revtex4-1}

\usepackage[english]{babel}
\usepackage{float}
\usepackage{amsfonts}
\usepackage{amssymb}
\usepackage{mathbrack}
\usepackage{siunitx}
\usepackage{mathtools}
\usepackage{bbold}
\usepackage{physics}
\usepackage{mwe}
\usepackage{amsthm,enumitem}
\usepackage{tikz}
\usepackage{ulem}
\usetikzlibrary{arrows,arrows.meta,shapes}
\usetikzlibrary{positioning}
\pdfoutput=1
\usepackage{dcolumn}
\usepackage{bm}
\usepackage[utf8]{inputenc}
\usepackage{graphics,graphicx}
\usepackage{parskip}
\usepackage{amsmath,amssymb}
\usepackage{multirow}
\usepackage{fancyhdr}
\usepackage{xcolor}
\usepackage{placeins}
\usepackage[title]{appendix}
\usepackage{wasysym}
\usepackage{url}
\usepackage{braket}
\usepackage{siunitx}
\usepackage{upgreek}

\begin{document}

\title{The helical quantum two-body problem and its wave packet dynamics}

\author{Peter Schmelcher}
\email{peter.schmelcher@uni-hamburg.de}
\affiliation{Zentrum f\"ur Optische Quantentechnologien, Fachbereich Physik,
Universit\"at Hamburg, Luruper Chaussee 149, 22761 Hamburg, Germany}

\date{\today}

\begin{abstract}
We explore the helical quantum two-body problem i.e. two repulsively Coulomb interacting particles
confined to move along a helix. The effective potential possesses a tunable number of potential wells
superimposed on the repulsive Coulomb interaction that can be varied by changing the ratio of the
pitch and radius of the helix. The anharmonicity of these wells depends crucially on this ratio and
on the order of the well which can be seen also by analyzing the individual wells energy eigenvalue 
spacing. Our main focus is the investigation of the quantum dynamics of differently prepared wave
packets that scatter from the multi-well potential landscape. We show that there exists a rich
pattern forming transient evolution which depends also on the number of bound states of the
individual wells. We demonstrate how the multiple wells leave their fingerprints
in the dynamics leading, among others, to oscillatory structures on different spatial scales, the formation of beats
and pulsed emission from single well localized wave packets due to their intrawell dynamics. 
\end{abstract}

\maketitle

\section{Introduction}
\label{intro}

\noindent
Helical structures appear in a variety of different biological, chemical and physical settings.
One of the most prominent examples is certainly the double helix of the deoxyribonucleic acid
due to the outstanding role this polymer plays for most organisms. Apart from DNA important
naturally occurring helical structures include $\alpha-$helices in proteins among many others. 
Here helical structures are often related to certain functionalities arising from the specific
geometry, such as biological information storage or cell proliferation \cite{Ilamaran19}.
A further example is the relationship between chirality of a molecule and its function \cite{Meierhenrich08}.
Apart from naturally occuring helices the controlled preparation and assembly of nano-scale
helices is an intense field of research \cite{Ren14,Shaikjee12}. Possible fields of applications of
nano-helices include, among others, optoelectronics \cite{Percec02}, sensors \cite{Yashima09},
responsive materials \cite{Pijper08}, logic devices \cite{Merkle96,Gupta23},
single molecule electronics \cite{Malyshev07}, and nano-scale machines \cite{Li14}.
Specifically helical carbon nanotubes for helix radii as small as $\approx 20 nm$ have been
prepared in the laboratory \cite{Ivanov94}. 
Helical structures are also encountered in the realm of cold and ultracold trapped ions and neutral atoms.
Harmonically confined ions can self-organize into intertwining helical arrays 
\cite{Hyde13,Landa13,Zade05,Tsytovich05,Birkl92,Bollinger94,Hasse90}. Ultracold neutral atoms can be trapped
in the evanescent light fields emerging from a nanoscale optical fibre \cite{Reitz12, Vetsch10,Vetsch12}.
Here the trapping light focus winds in a helical manner around the fiber and can serve as a waveguide
for either neutral atoms or dipolarly interacting atoms/molecules.

\noindent
The above-described helical configurations have been one of the motivations for the development of 
a helical charged particle model. In this model the charged particles are confined/constrained 
to move along a one-dimensional helix while interacting via the dynamically forbidden
three-dimensional space. This model has been shown to exhibit several
unusual and interesting properties both in terms of the resulting structures and dynamical behaviour.
First of all the effective two-body interaction among the particles appears to be of oscillatory character
and the purely repulsive character of the 3D Coulomb interaction can now develop local potential wells
\cite{Kibis92,Schmelcher11}. The number of those wells is finite and depends on the pitch and radius
of the helical confinement. For two particles the stable equilibria are formed by positioning the particles
(approximately) at opposite sides of the winding of the helix, i.e. they form a displaced configuration.
Increasing the number of equilibria and number of particles rapidly leads to a very complex many-body
potential landscape which 'lives' between regularity and randomness thereby showing clustering behaviour
\cite{Doerre25} evolving with increasing particle number. 

\noindent
Several investigations of the helical few- to many-body
behaviour have been performed in the past decade. In ref.\cite{Zampetaki15-1} the evolution of the
classical band structure on a toroidal helix with varying helical radius has been explored thereby unraveling
a route to degeneracy and band inversion followed by a bifurcation from a single to two-band structure
\cite{Zampetaki15-1}. The extension to the regime of nonlinear excitation dynamics \cite{Zampetaki15-2}
demonstrated a crossover from dispersion and revival dynamics in the lattice to 'fragmentation'
into breatherlike excitations. Adding external forcing and dissipation the existence of a structural
crossover and a pinned-to-sliding transition \cite{Zampetaki17} has been found. The electrostatic
bending response of a chain of particles on a finite helical filament shows a rich anomalous
behaviour depending on the geometry parameters, the filling and the size \cite{Zampetaki18}. 
In ref.\cite{Plettenberg17} the impact of an external electric field on the equilibria has been
addressed and it was shown how the field can be used for state transfer. More recently it has
been demonstrated \cite{Siemens20} that the electric field can also be employed to trigger a 
crossover from amorphous to crystalline order in the helical particle configurations.
Driving the system with a time-dependent external field allows to split the chaotic phase
space region which has a strong impact on the direction of the resulting transport \cite{Siemens21}
Augmenting this setup with a static field one could show that the single particle physics represents
that of a generalization of the Kapitza pendulum \cite{Gloy22}.

\noindent
While all of these investigations of helically confined Coulomb interacting particles
are on their classical properties and dynamics there is no ab initio investigation of the corresponding quantum
properties and dynamics up to date. It is only for dipolarly anisotropically interacting particles
that there exist a very few works on their bound states \cite{Pedersen14,Pedersen16-1,Pedersen16-2}.
The present work aims at performing a first step in order to address this gap by exploring the helical quantum two-body problem. 
After establishing the quantum two-body Hamiltonian we will analyze the properties of the underlying effective
potential ${\cal{V}}$ thereby addressing the number and shape of individual potential wells occuring in ${\cal{V}}$.
This includes a concise inspection of the spectrum of bound states for these (isolated) individual wells which
appear to be only metastable states with a finite lifetime as they occur in ${\cal{V}}$. The main results
of this work concern the wave packet (WP) quantum dynamics of the two-body problem taking place in the highly
oscillatory helical potential landscape. We explore the time evolution of the density of the wave packets for different
helical potentials with varying number of potential wells and demonstrate its dependence on the placement
of the initial WP. Our focus is hereby on the intermediate time evolution and not on the asymptotic
$t \rightarrow \infty$ behaviour. The initial Gaussian undergoes a very rich patternforming interference dynamics, including,
among others, pulses with multiple beats or a regular to irregular sequence of peaks. A rich intrawell vs.
interwell dynamics is unraveled leading to the interplay of fragmentation, above barrier and tunneling dynamics. The dynamics
is also analyzed by exploring the time evolution of the occupation of the different individual wells as well as expectation
values and mean square displacements of the intrawell dynamics.

\noindent
We proceed as follows. In section \ref{sah} we introduce our setup and Hamiltonian and discuss
the properties of the oscillatory helical potential. In section \ref{iwpas} the shape of individuell
wells and their spectral properties are discussed. Section \ref{qdwp} contains an analysis of the density
evolution of WPs with different initial conditions and for two different Hamiltonian. We also
explore the time evolution of the resulting occupation probability of individual wells and the
corresponding intrawell dynamics. Finally, in section \ref{concl} we provide our conclusions and
outlook.

\section{Setup and Hamiltonian}
\label{sah}

\begin{figure}[H]
\hspace*{-3cm}
\includegraphics[width=12cm,height=8cm]{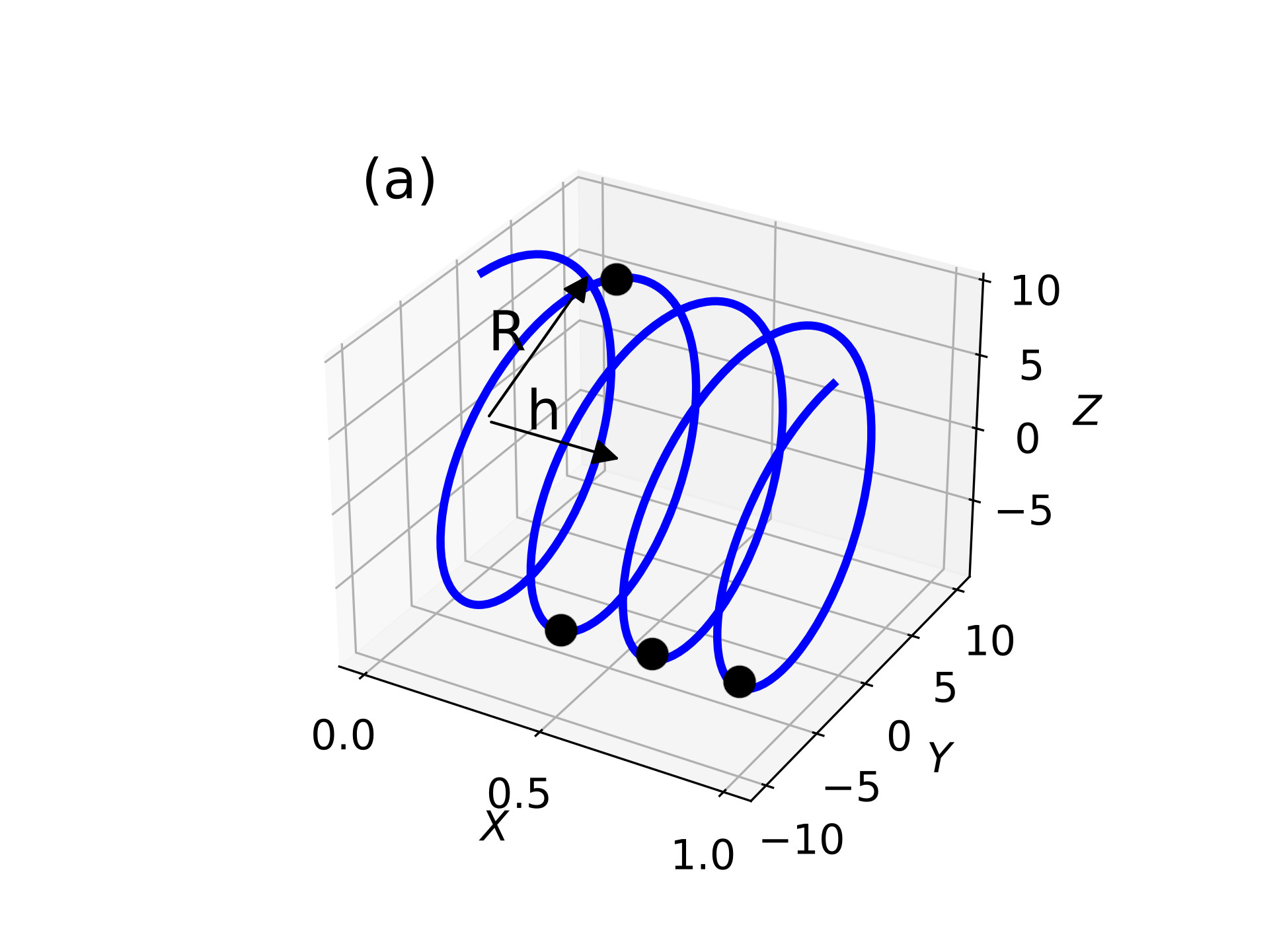}
\hspace*{-1.5cm}
\includegraphics[width=10cm,height=7cm]{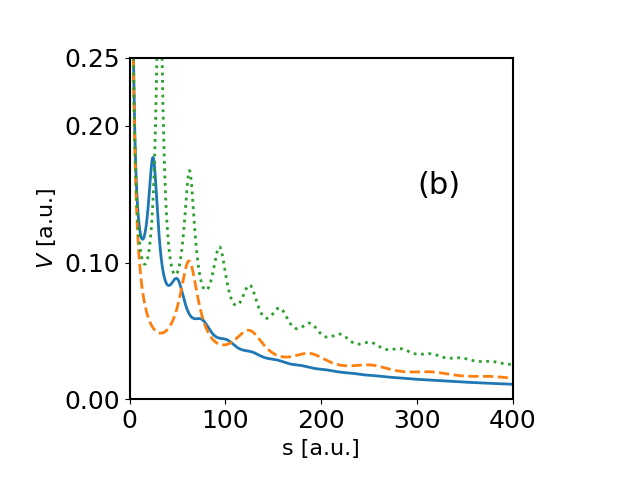}
\vspace*{-0.5cm}
\caption{(a) Sketch of a helix with the radius $R$ and pitch $h$. The black spheres
on the helix indicate equilibria of two particle positions: if one particle is on
the top of the winding the other is approximately at the bottom of the next or all
following windings. Whether such a configuration represents a minimum and consequently
forms a potential well depends on the detailed values of the radius and pitch, see text.
(b) The helical potential $V$ as a function of the path length $s$ for three cases:
$h=5.8,R=4$ (solid line, three potential wells), $h=10,R=10$ (dashed line, six potential
wells) and $h=3,R=5$ (dotted line, 17 potential wells, not all shown).}
\label{Fig:1}
\end{figure}

\noindent
We assume our charged particles to be confined to a strictly one-dimensional helix with constant pitch $h$ and
radius $R$, see the sketch of Fig.\ref{Fig:1}(a). The helix is given by the locus of all points
$(x,y,z)= (R \cdot \text{cos} \left( \frac{2\pi t}{h} \right), R \cdot \text{sin} \left( \frac{2 \pi t}{h} \right), t)$
where e.g. $t \in \left[0,h\right]$ describes one winding. It possesses a constant curvature and torsion.
We parametrize the helix by its path length 
$s = \sqrt{1 + \left( \frac{2 \pi R}{h} \right)^2 } t$. The single particle kinetic energy on the strictly
one-dimensional helix can be obtained from that of an infinitely thin waveguide, see refs.\cite{Stockhofe14,Costa81,Araujo26a},
and appears to be in the canonical form $\frac{1}{2M} p_k^2$ with $p = \frac{1}{i} \frac{\partial}{\partial s_k}$,
where $k$ is the particle index. We assume $\hbar = e = 1$ in the following, where $e$ is the elementary charge.

\noindent
Placing two equally charged particles with the same mass $M_1=M_2=M$ on the helix leads to the two-body 
Hamiltonian 

\begin{equation}
{\cal{H}} = \frac{1}{2M} p_1^2 + \frac{1}{2M} p_2^2 + {\cal{V}} \left(s_1,s_2\right)
\label{eq:1}
\end{equation}

\noindent
We now exploit the fact that the helix is the only curved manifold that shows a separation
of the center of mass and relative motion for an interacting particle system \cite{Zampetaki13}.
Transforming from the path length coordinates and momenta $s_i, p_i, i=1,2$ to the corresponding relative
and center of mass coordinates $s = s_1-s_2, S = \frac{s_1+s_2}{2}$ yields for the Hamiltonian

\begin{equation}
{\cal{H}} = \frac{1}{4M} p_S^2 + \frac{1}{M} p_s^2 + {\cal{V}} \left( s \right)
\label{eq:2}
\end{equation}

\noindent
The CM separates and the corresponding kinetic energy can therefore be neglected from here on.
For the relative motion we therefore arrive at our working Hamiltonian

\begin{equation}
{\cal{H}}_s = \frac{1}{M} p_s^2 + {\cal{V}} \left( s \right)
\label{eq:3}
\end{equation}

with

\begin{equation}
{\cal{V}} = \left( 
       2 R^2 \left(1 - \cos\left(\frac{s}{\beta} \right) \right) + \left( \frac{h}{2 \pi \beta} \right)^2 \cdot s^2 
       \right)^{-\frac{1}{2}}
\label{eq:4}
\end{equation}

\noindent
with $\beta = \sqrt{(\frac{h}{2 \pi})^2 + R^2}$. The helical potential ${\cal{V}}$ possesses, besides
the Coulombic term $\propto s^2$, an oscillatory contribution $\propto \text{cos} \left( \frac{s}{\beta} \right)$
with a spatial frequency $\frac{1}{\beta}$
which, for certain parameter domains, leads to the existence of a number of local potential wells \cite{Schmelcher11}
as we shall summarize in the following. As a consequence of the helical confinement the purely repulsive 3D Coulomb
interaction therefore turns into an oscillatory potential landscape. The physical picture behind this behaviour
is the fact that the two-particle interaction energy is sufficiently lowered while 
placing the charges on opposites sides of corresponding windings whereas it is raised if they are on the same
sides of the windings, see the sketch in Fig.\ref{Fig:1}(a). Fig.\ref{Fig:1}(b) shows the helical potentials
for the cases $(h=5.8,R=4)$, $(h=10,R=10)$ and $(h=3,R=5)$. These potentials possess three, six and seventeen potential
wells, respectively. Their depths decrease with increasing order, i.e. increasing value of $s$ for $s>0$.
Each minimum corresponds classically to a stable equilibrium configuration that is accompanied by a
phase space region of librational motion. Note that minima (see Fig.\ref{Fig:1}(a)) correspond
to configurations where the two particles are on opposite sides of either the same or different
windings whereas maxima mean that they are on the same side of different windings.
Let us derive the criterion for the existence of extrema. 
$\frac{d{\cal{V}}}{ds}=0$ yields 

\begin{equation}
        \sin\left(\frac{s}{\beta} \right) = - \left( \frac{h^2}{4 \pi^2 R^2 \beta} \right) s
        \label{eq:5}
\end{equation}

\noindent
For given $(h,R)$ this implicit equation provides with a set of extrema values for $s$ by 
determining the intersection points of the straight line with the sine function.
The point of the emergence of new extrema occurs for the case of the same derivative of the 
sine and the linear function. This leads to the relationship

\begin{equation}
        s = \beta \sqrt{ \left(\frac{4 \pi^2 R^2}{h^2}\right)^2 - 1}
\label{eq:6}
\end{equation}

\noindent
and resultingly an implicite equation for the helix parameter values at which extrema emerge

\begin{equation}
        \cos \left( \sqrt{ \left(\frac{4 \pi^2 R^2}{h^2}\right)^2 - 1} \right) = - \frac{h^2}{4 \pi^2 R^2}
\label{eq:7}
\end{equation}

\noindent
Note that this relation depends only on the ratio $r = \frac{h}{R}$ and not on the values of the pitch
and radius separately. The first few values of this ratio (bifurcation points) corresponding to the
existence of one to five potential wells read $r \approx 2.929, 1.899, 1.513, 1.295, 1.151$. 
For $r \rightarrow 0$ we obtain approximately $\cos \left( \frac{4 \pi^2}{r^2} \right) = 0$. This yields
$r = \sqrt{\frac{8 \pi}{2n+1}}$. 

\noindent
We note that performing a canonical scale transformation $s^{\prime} = \frac{s}{\beta}, p^{\prime}_s = \beta p_s$ 
yields the transformed Hamiltonian

\begin{equation}
        H^{\prime}_s \cdot M \beta^2 = {p^{\prime}_s}^2 +  \frac{2 \pi M \beta^2}{h} \left( 
        8 \pi^2  \frac{R^2}{h^2} \left(1 - \cos\left(s^{\prime} \right) \right) 
        + s^{\prime 2} \right)^{-\frac{1}{2}}
\end{equation}

\noindent
This shows that after performing the canonical scaling the potential term depends, apart from
a prefactor $\frac{2 \pi M \beta^2}{h}$ only on the ratio $\frac{h}{R}$ and not on $h$ and $R$
separately, with importance for the following discussion of individual wells properties.

\section{Individual well properties and spectra}
\label{iwpas}

\begin{figure}[H]
\centering
\includegraphics[width=10cm,height=7cm]{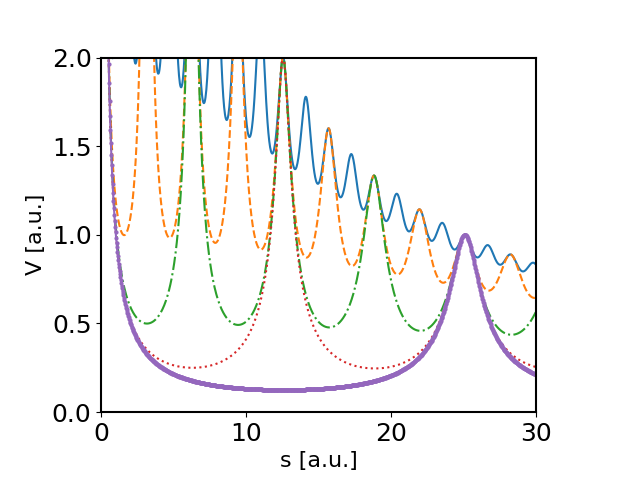}
\caption{A zoom into a specific window of the helical potential ${\cal{V}}$ as a function of the
path length $s$ for $h/R=0.25$ for $R=0.25,0.5,1.0,2.0,4.0$ corresponding to the solid, dashed, dot-dashed,
dotted and solid curves from top to bottom, in order to illustrate the appearance/shape
of the individual potential wells.}
\label{Fig:1A}
\end{figure}

\noindent
It is instructive to analyze the properties of individual wells of the potential ${\cal{V}}$. For that
purpose Fig.\ref{Fig:1A} shows a specific region of ${\cal{V}}(s)$ for $h/R=0.25$ and different absolute
values of $(h,R)$. 
We observe that the extension of the potential wells changes from narrow to broad when increasing the
values for both the pitch and the radius. Correspondingly the positions of the minima increase with increasing
helical parameter values. As analyzed above decreasing $\frac{h}{R}$ means increasing the number of potential wells
for a given interval of $s$. The individual wells are strongly anharmonic and it can be shown that this
strong anharmonicity is maintained for changing $(h,R)$ while keeping their ratio fixed, see Fig.\ref{Fig:1A}
and in particular the scaling analysis of the previous section. Changing the ratio $\frac{h}{R}$ from values
smaller than one to values larger than one decreases the degree of anharmonicity, as a corresponding
quantitative analysis shows, and the harmonic approximation becomes increasingly valid. One should 
however keep in mind, that individual wells are never parity symmetric around their minimum such that there is always
a systematic deviation from a harmonic approximation. For given values of $h$ and $R$ the anharmonicity
increases with the order of the individual well i.e. outer wells are systematically more anharmonic 
as compared to inner wells. At the same time the depth of the wells decreases monotonically with their order:
the deepest (shallowest) wells are the innermost (outermost) ones.

\noindent
In the spirit of our analysis of individual wells, let us now inspect exemplarily the spectral properties
of individual isolated wells. We herefore cut out individual wells from the helical potential ${\cal{V}}$
by setting the value of the potential ${\cal{V}}$ constant to the left and right of the 
position $s$ where it reaches the value of the energy maximum of the corresponding well. 
This way the quasibound states of the wells of ${\cal{V}}$ become strictly
bound states. We compute the spectra using a $10-$th order finite difference discretization \cite{Groenenboom90}
and a corresponding diagonalization of the discretized Hamiltonian providing us with
the eigenvalues and eigenvectors of the individual well problem.

\begin{figure}[H]
\centering
\includegraphics[width=7cm,height=6cm]{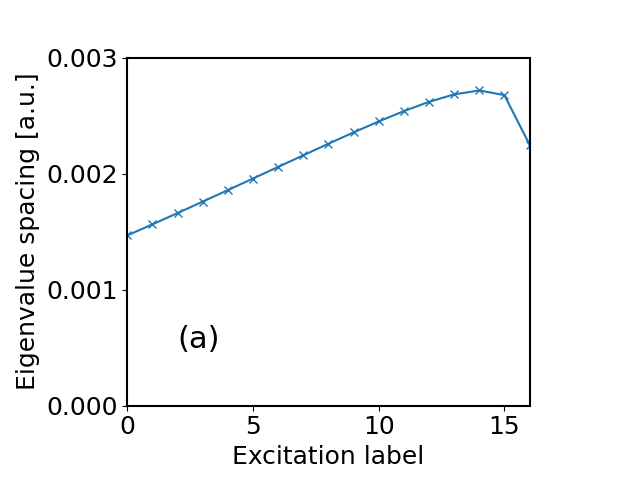}
\includegraphics[width=7cm,height=6cm]{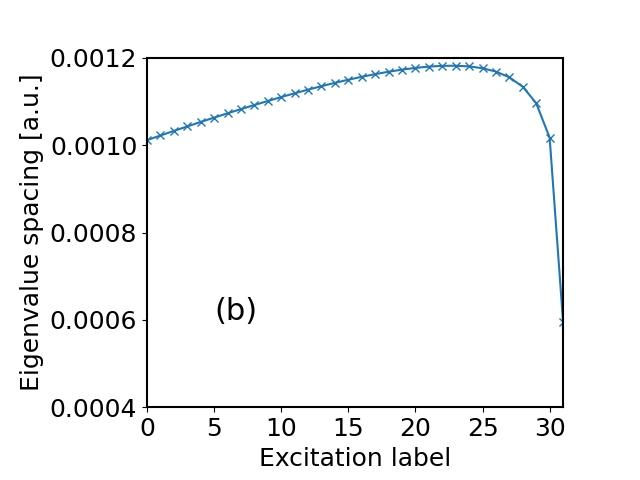}
\caption{(a) Energy eigenvalue spacing in the isolated innermost well of the helical potential
${\cal{V}}$ for $h=20,R=20,M=1$. The eigenvalue spacing increases linearly for the $18$ bound
states except at the threshold to unboundedness. (b) the same but for the parameter values
with $h=12.2,R=10,M=100$ where $32$ bound states exist. The linear slope is here much smaller,
indicating the approximate equidistant spectrum.}
\label{Fig:1B}
\end{figure}

\noindent
Fig.\ref{Fig:1B}(a) shows the spacing of the discrete energy eigenvalues
for the isolated innermost well of the helical potential ${\cal{V}}$ for $h=20,R=20$ and $M=1$.
The eigenvalue spacing increases approximately linearly with increasing degree of excitation for the first $13$ spacings
and consequently bends and starts to decrease for the very few states close to the potential threshold 
passing to the asymptotically constant behaviour. We encounter approximately an increase of $100 \%$ of the spacing which
indicates that the spectrum belongs to a strongly anharmonic potential well.
Fig.\ref{Fig:1B}(b) shows the eigenvalue spacing for the innermost well of the potential ${\cal{V}}$
for $h=12.2,R=10$ and $M=100$ where $32$ bound states exist. The linear slope is here much smaller,
and the increase of the spacing is approximately $20 \%$ - this indicates that the anharmonicity
of the underlying potential well is much weaker compared to the case of Fig.\ref{Fig:1B}(a).

\section{Quantum dynamics and wave packet evolution}
\label{qdwp}

\noindent
This section is dedicated to the analysis of the quantum dynamics in the helical potential by employing
a Gaussian WP that is initially placed at different positions in the potential. 
Our goal is to demonstrate the variability in the dynamics due to interference, above barrier and tunneling effects for
different potentials and by using suitable analysis tools such as the individual well occupation (IWO) or
intrawell expectation values. We emphasize that we do not address the asymptotic $t \rightarrow \infty$ scattering properties
but are focusing on the intermediate time quantum dynamical behaviour which illuminates the relevant
processes taking place in the oscillatory potential surface.

\subsection{Quantum dynamics in the helical landscape for $h=5.8,R=4,M=1$}
\label{qdcase1}

\noindent
The helical potential for $h=5.8, R = 4.0$ possesses three local potential
wells with minima at positions $s \approx 13, 41, 69$. The (isolated) inner well is deep enough to accomodate for
a single bound state, whereas the second and third well possess no bound states, the latter being
very shallow. We focus on the case $M=1$ and cut off the helical potential sufficiently close to the singularity,
such that the impact on all following results is marginal. Although originally designed for integrating
the dynamics of many distinguishable (vibrational) modes we use here the Multi-Configuration Time-Dependent Hartree Method
(MCTDH), as well-documented in refs.\cite{Meyer90,Manthe92,Beck00,Meyer09}, for integrating
our underlying time-dependent Schr\"odinger equation. As a fundamental grid method we use the sine-DVR with the typical 
number of grid points ranging between $2 \cdot 10^3$ and $4 \cdot 10^3$ thereby covering the coordinate
$s-$space such that the WP is well-separated from the boundaries of the grid. As a result
we can work with open boundary conditions and no complex absorbing potentials are necessary.

\begin{figure}[H]
\centering
\includegraphics[width=8cm,height=6cm]{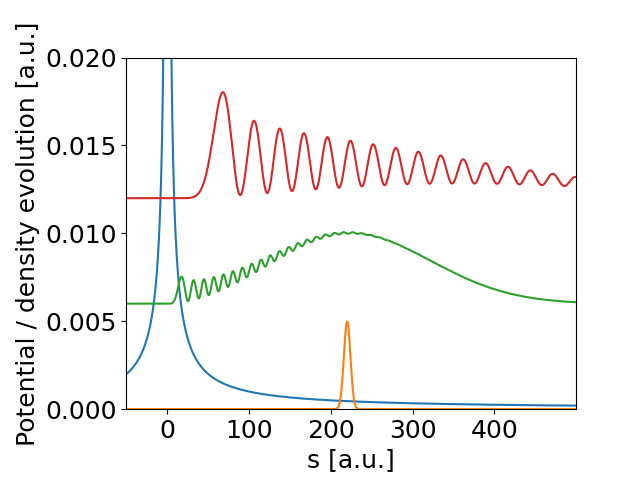}
\caption{The Coulomb potential and snapshots of the density time evolution
for times $t=0,400,1000$ (from bottom to top) for an initial Gaussian wave packet located at $s_0 = 220.0$
with zero momentum and width $ \Delta s = 4.0$ for $M=1$. The underlying sine DVR grid has 2301 points in the interval
$[-150,1000]$. Note that both the potential and the densities are scaled (and shifted)
to fit the same coordinate scale.}
\label{Fig:2}
\end{figure}

\noindent
Our initial ($t=0$) WP reads

\begin{equation}
\Psi(s,t=0) = {\cal{N}} {\text{exp}}\left(-\frac{1}{4} \left(\frac{\left(s-s_0 \right)}{\Delta s} \right)^2 \right)
{\text{exp}} \left(i p_0 \left(s-s_0 \right) \right)
\end{equation}

\noindent
with the normalization ${\cal{N}} = \left(2 \pi \left(\Delta s \right)^2 \right)^{-\frac{1}{4}}$, 
the maximum position $s_0$, the width $\Delta s$ and the momentum $p_0$. To set the stage and for
reasons of comparison we remind the reader of what happens in the case of a pure Coulomb potential. Fig.\ref{Fig:2}
shows three snapshots of the density of the wave packet (DWP) for $t=0,400,10^3$ from bottom to top for a 
WP with $s_0=220, \Delta s = 4.0, p_0=0$. At $t=400$, due to reflection at the repulsive Coulomb
potential, the onset of an oscillatory behaviour on top of the left branch of the Gaussian envelope is observed.
At $t=10^3$ the DWP has largely reshaped into a backreflected oscillatory pulse with a monotonically
decaying envelope structure. The approximately constant wave length of the oscillations is 
determined by the magnitude of the momentum of the initial WP. With increasing time this pulse roles 
down the Coulomb potential towards larger values of $s$ while maintaining its shape during time evolution 
but increasing the wavelength and decreasing the amplitude of its spatial oscillations.

\begin{figure}[H]
\centering
\includegraphics[width=8cm,height=6cm]{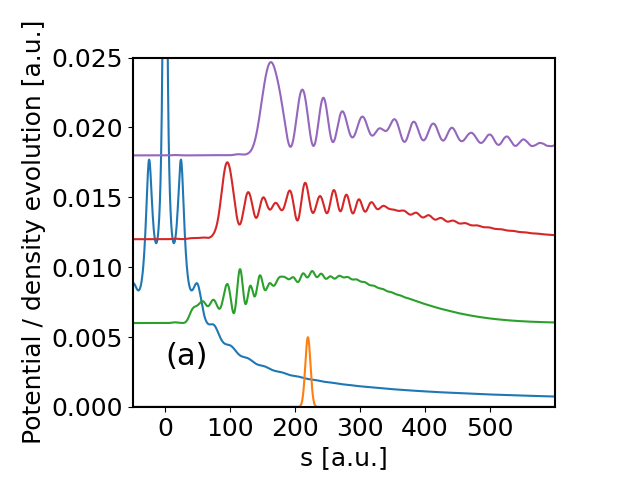}
\includegraphics[width=8cm,height=6cm]{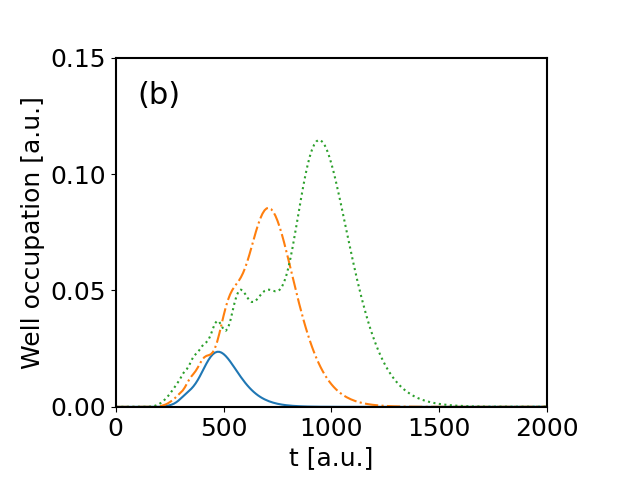}
\caption{(a) The helical potential (h=5.8,R=4.0) possessing three distinct minima and snapshots of the density time evolution
for times $t=0,500,700,1000$ (from bottom to top) for an initial Gaussian wave packet located at $s_0 = 220.0$
with zero momentum and width $ \Delta s = 4.0$ for $M=1$. The underlying sine DVR grid has 2301 points in the interval
$[-150,1000]$. Note that both the potential and the densities are scaled (and shifted)
to fit the same coordinate scale. 
(b) The evolution of the density integrated over the first innermost (solid line), second (dash-dotted line) and 
third outermost (dotted line) well of the potential.}
\label{Fig:3}
\end{figure}

\noindent
We use again the above initial WP but now expose it to the helical potential for $h=5.8,R=4$. Our choice
$s_0 = 220$ means that the initial WP is localized practically completely outside the regime of the potential
wells. It should be noted, that all below given density snapshots are, for each time evolution, scaled with
different scaling factors in order to fit into a single figure. Our first focus is on zero momentum initial WP to probe the 
helical potential landscape. Fig.\ref{Fig:3}(a) shows the snapshots of the DWP for $t=0,500,700,10^3$. 
Upon the spreading of the left branch of the originally Gaussian WP towards the helical potential and its wells
we observe the formation of an oscillatory structure, like in the above pure Coulomb case, but now it
exhibits a different substructure: we observe an envelope behaviour with a nodal structure or, in other
words, the emergence of a few beats in the developing density pulse. This pulse travels for longer times to larger
distances while maintaining its overall structure but changing the corresponding amplitudes.

\noindent
Fig.\ref{Fig:3}(b) shows the corresponding time evolution of the populations or occupation,
i.e. the integrated densities, between
two maxima i.e. each one covering an individual well (IWO) of the helical potential. As expected the increase of the
IWO with time happens first for the outermost followed by the inner wells. The innermost well
possesses the smallest IWO which decays at much earlier times compared to the two outer ones in the
course of the dynamical WP reflection. The outermost well possesses the largest IWO for the longest
time span. All IWOs exhibit a single dominant and broad peak but the IWO of the outermost well 
presents an additional pronounced shoulder for shorter times with small amplitude oscillations on top. 
Comparing to the above-discussed Coulombic WP scattering, it is therefore evident already
for this first study case, that the presence of the wells in the helical potential leaves its 
fingerprints in the interference dynamics of the WP.

\noindent
We now address the dynamical evolution of a WP that initially is centered at the minimum of the
second to inner well i.e. $s_0 = 40.99$ and possesses a width of $\Delta s = 6$. We remind the
reader of the fact that this well possesses no (quasi-) bound states.

\begin{figure}[H]
\centering
\includegraphics[width=8cm,height=6cm]{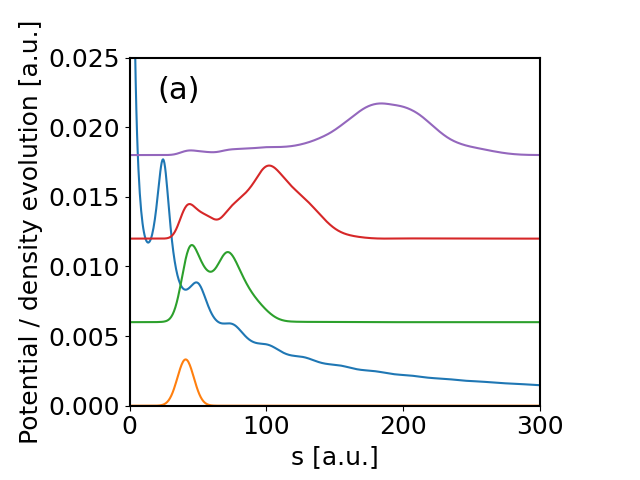}
\includegraphics[width=8cm,height=6cm]{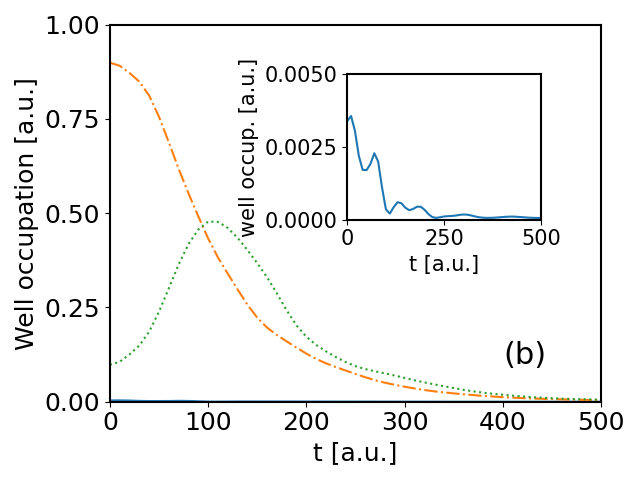}
\caption{(a) The helical potential (h=5.8,R=4.0) possessing three distinct minima and snapshots of the density time evolution
for times $t=0,140,230,400$ (from bottom to top) for an initial Gaussian wave packet located at $s_0 = 40.99$
with zero momentum and width $ \Delta s = 6.0$ and $M=1$. The underlying sine DVR grid has 2301 points in the interval
$[-150,1000]$. Note that both the potential and the densities are scaled (and shifted)
to fit the same coordinate scale. 
(b) The evolution of the density integrated over the first innermost (solid line), second (dash-dotted line) and 
third outermost (dotted line) well of the potential. The inset shows a zoom into the evolution of the integrated
density for the first well.}
\label{Fig:4}
\end{figure}

\noindent
Fig.\ref{Fig:4}(a) shows the evolution of the DWP on the basis of snapshots for $t = 0,140,230,400,10^3$.
At intermediate times the originally single-humped DWP develops a broad double peak structure
where the first inner peak is roughly localized in the second inner well of the potential and
the second peak is localized around the following third well, see the snapshot for $t=140$. With increasing evolution time
the first peak decays and the second peak moves to larger distances, see the snapshot for 
$t= 230$. At $t= 400,1000$ only an inner shallow tail of the density remains
while its single asymmetric broad peak moves outward to larger distances $s$.
Fig.\ref{Fig:4}(b) shows the corresponding IWO for the three wells.
While the innermost well is only very weakly populated for short times (see inset), the
second well possesses by preparation of the WP a dominant large IWO for $t=0$ which 
decays monotonically with increasing time. The third outermost well has an approximately
ten percent IWO at $t=0$ and increases thereafter.

\begin{figure}[H]
\centering
\includegraphics[width=8cm,height=4cm]{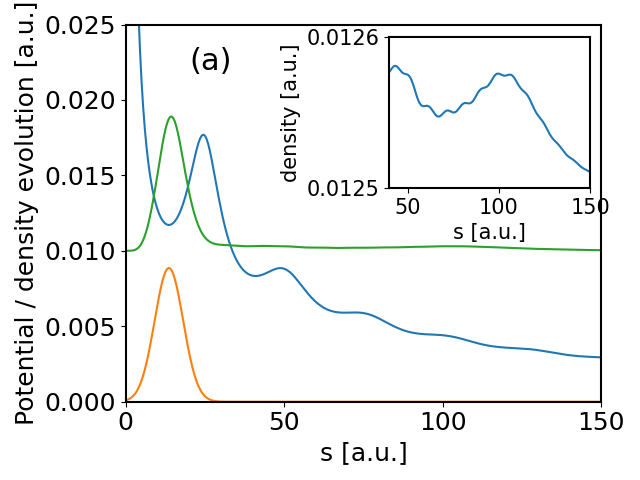}
\includegraphics[width=8cm,height=4cm]{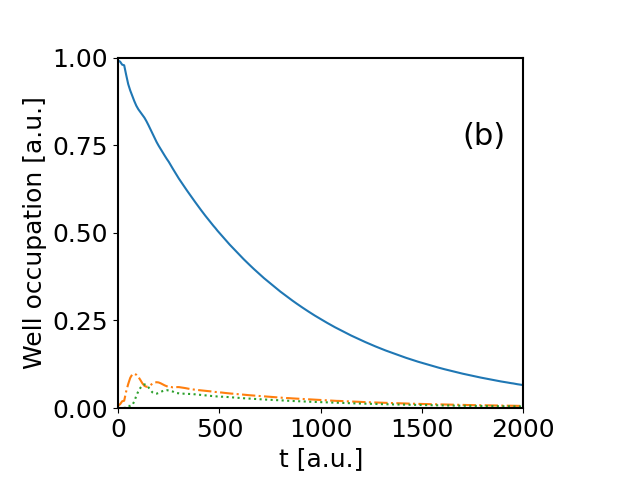}
\caption{(a) The helical potential (h=5.8,R=4.0) possessing three distinct minima and snapshots of the density time evolution
for times $t=0,200$ (from bottom to top) for an initial Gaussian wave packet located at $s_0 = 13.63$
with zero momentum and width $ \Delta s = 4.5$ for $M=1$. The underlying sine DVR grid has 3301 points in the interval
$[-150,1500]$. Note that both the potential and the densities are scaled (and shifted)
to fit the same coordinate scale. The inset shows a zoom into the tail of the density distribution for $t=200$.
(b) The evolution of the density integrated over the first innermost (solid line), second (dash-dotted line) and 
third outermost (dotted line) well of the potential.}
\label{Fig:5}
\end{figure}

\noindent
At $t \approx 100$ the second and third well possess the same IWO
which represents (approximately) also the maximum probability in the time evolution of the third well. 
For large times $t > 300$ both IWOs became very small and are further decreasing according
to a similar tail-like evolution. Obviously this behaviour is very distinct from the one shown 
above for the case $s_0 =220$.
As a natural next step we explore the time evolution for an initial Gaussian WP
centered at $s_0 = 13.63$ i.e. in the innermost well of the potential, for zero momentum, $M=1$ and width $\Delta s = 4.5$.
Fig.\ref{Fig:5}(a) shows the corresponding snapshots for times $t = 0, 200$. The inset shows a zoom into the
tail of the density distribution for $t = 200$. For such comparatively short times there is a weak
leaking of probability from the innermost well to large distances $s$. For longer times (see
Fig.\ref{Fig:5}(b) up to $t=2000$) there is a strong monotonous decrease of the inner well IWO
such that for $t=2000$ only $10 \%$ of the original probability is maintained. The small higher order wells
IWO becomes largest for short times and decays thereafter very slowly on the mentioned large time scale.

\begin{figure}[H]
\centering
\includegraphics[width=8cm,height=4cm]{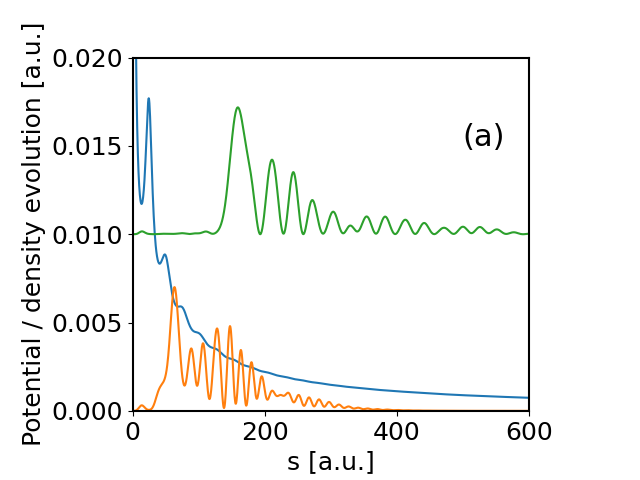}
\includegraphics[width=8cm,height=4cm]{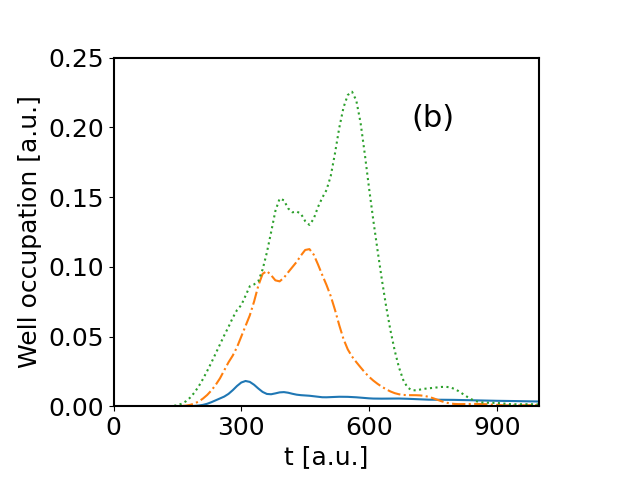}
\caption{(a) The helical potential (h=5.8,R=4.0) possessing three distinct minima and snapshots of the density time evolution
for times $t=550,1000$ (from bottom to top) for an initial Gaussian wave packet located at $s_0 = 220.0$
with momentum $p_0 = - 0.154$ ($M=1$) and width $ \Delta s = 4.5$. The underlying sine DVR grid has 3301 points in the interval
$[-150,1500]$. Note that both the potential and the densities are scaled (and shifted)
to fit the same coordinate scale. 
(b) The evolution of the density integrated over the first innermost (solid line), second (dash-dotted line) and 
third outermost (dotted line) well of the potential.}
\label{Fig:6}
\end{figure}

\noindent
Let us now consider the case of a finite (negative) momentum $p_0 = - 0.154$ 
and relocating the initial WP back to $s_0 = 220$ i.e. far outside the region of the potential wells.
Fig.\ref{Fig:6}(a) shows the snapshots for the times $t = 550, 1000$. The scattering off the potential
wells produces for intermediate times a DWP consisting of a main peak followed by two beats significantly
different in amplitude, see snapshot for $t=550$. For $t=1000$ the beats are still present but the DWP 
has overall moved to larger distances. The IWOs, see Fig.\ref{Fig:6}(b), show therefore
substantial nonzero values for all three wells only for $200 < t < 700$ with a similar structure to
the one shown above.

\begin{figure}[H]
\centering
\includegraphics[width=8cm,height=6cm]{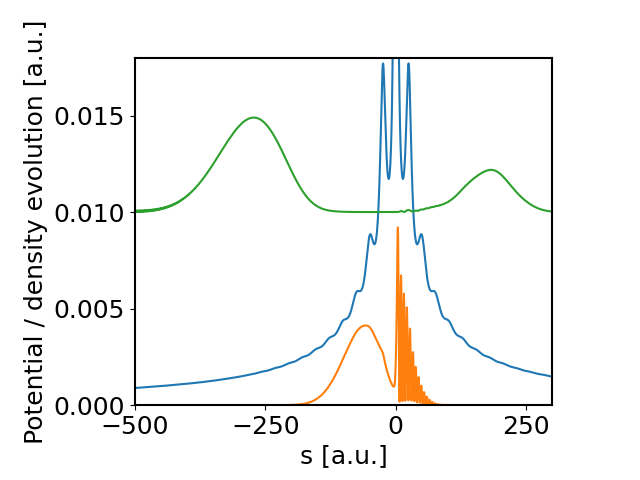}
\caption{The helical potential (h=5.8,R=4.0) possessing three distinct minima and snapshots of the density time evolution
for times $t=180,310$ (from bottom to top) for an initial Gaussian wave packet located at $s_0 = 220.0$
with momentum $p_0 = - 0.8$ and width $ \Delta s = 4.5$ for $M=1$. The underlying sine DVR grid has 4001 points in the interval
$[-500,1500]$. Note that both the potential and the densities are scaled (and shifted)
to fit the same coordinate scale.}
\label{Fig:7}
\end{figure}

\noindent
Finally, for the same helical potential (h=5.8,R=4.0) we choose a much higher energy for the
WP by imposing the parameter values $p_0 = -0.8, s_0=220.0, \Delta s=4.5$. Fig.\ref{Fig:7}
shows two snapshots at times $t= 180,310$. The WP possesses now sufficient energy to surpass
the (regularized) Coulomb singularity. The time instant $t=180$ shows an intermediate stage
of the transmission process where the single broad peak of the transmitted part of the DWP
is emerging while the backreflected part shows (due to the high energy) many short wavelength
oscillations that accumulate for small values of $s$. At $t=310$ the transmitted and reflected
WP are spatially well-separated and each one shows a single broad peak structure.

\subsection{Quantum dynamics in the helical landscape for $h=10,R=10,M=1$}
\label{qdcase2}

\noindent
This subsection is dedicated to the helical landscape for pitch and radius being $h=10, R=10$.
It possesses six potential wells with minima at positions $s \approx 32.5, 98, 163, 229, 295, 361$.
The innermost (isolated) well is deep enough to accomodate for
five bound states, the second well for three states and the third as well as fourth well for a single state.
All further outer wells have no bound states. Due to the existence of the larger number of wells
as well as their few bound states we expect a different phenomenology of the WP dynamics compared to the previous
case $h=5.8,R=4.0$.

\begin{figure}[H]
\centering
\includegraphics[width=8cm,height=6cm]{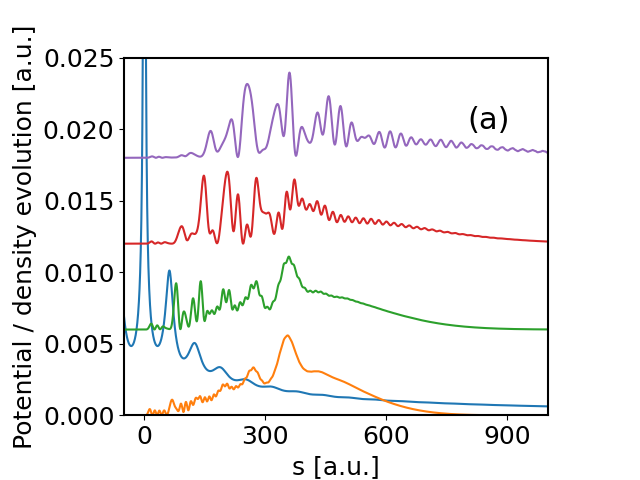}
\includegraphics[width=8cm,height=6cm]{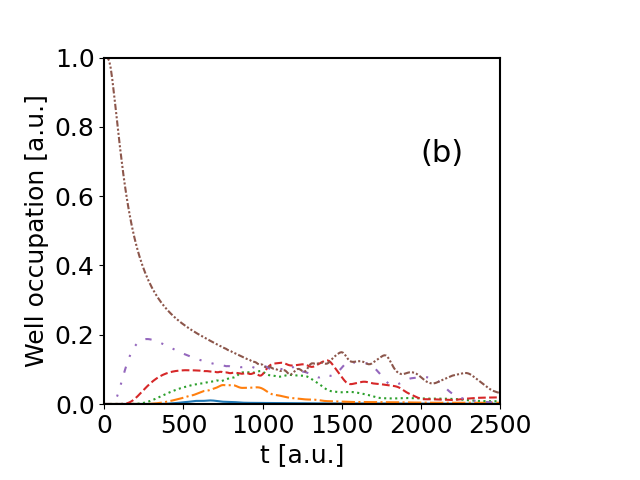}
\caption{(a) The helical potential (h=10,R=10) possessing six distinct minima and snapshots of the density time evolution
for times $t=600,750,1100,1500$ (from bottom to top) for an initial Gaussian wave packet located at $s_0 = 350.0$
with zero momentum and width $ \Delta s = 4$ for $M=1$. The underlying sine DVR grid has 4301 points in the interval
$[-150,2000]$. Note that both the potential and the densities are scaled (and shifted)
to fit the same coordinate scale. 
(b) The evolution of the density integrated separately and individually
over the first innermost to the sixth outermost well of the potential.
The linestyles for the first to the sixth integrated well probabilities are solid, dash dotted, dotted, dashed, dash dot dotted 
and densely dash dot dotted.}
\label{Fig:8}
\end{figure}

\noindent
We start the study of the DWP evolution with an initial WP located at $s_0 = 350$, i.e. close 
to the minimum of the sixth well, for a width
of $\Delta s = 4$ and zero momentum $p_0 = 0$. Fig.\ref{Fig:8}(a) shows the snapshots of the
evolution for the times $t=600,750,1100,1500$ and it becomes immediately evident that there
is a more complex pattern formation in the DWP evolution. 
At $t = 600$ the density has still a main peak at $s_0 = 350$ but a very different shape
for shorter as compared to large distances. For shorter distances there is several oscillations
equipped with smaller amplitude higher frequency suboscillations. For larger distances
the spreading of the DWP is smooth and monotonically decreasing: it does encounter neither the
sequence of individual wells nor the (strongly) repulsive Coulomb branch. With further evolution time
(see snapshots for $t=750,1100$ of Fig.\ref{Fig:8}(a)) the inner shorter distance 
behaviour of the DWP is restructured and passes through a period of reshaping the peaks
thereby encompassing a higher frequency content. At $t=1000$ three large main peaks
are established followed, for larger distances, by a decaying tail with a regular oscillatory structure on top
of it. For $t=1500$ the DWP has developed into a pulse that shows, from short to large
distances, a sequence of peaks with increasing height followed by a sequence of beats
with decreasing amplitude. This structure travels with increasing time as a whole towards
larger distances. Fig.\ref{Fig:8}(b) shows the IWOs which, except for the innermost well
IWO, all participate substantially in the scattering process. However, due to preparation,
for short times the IWO of the sixth well is dominant but decreases rapidly with
increasing time. For long times the IWO decrease step by step with their corresponding 
order.

\begin{figure*}[htpb!]
\centering
\includegraphics[width=\textwidth]{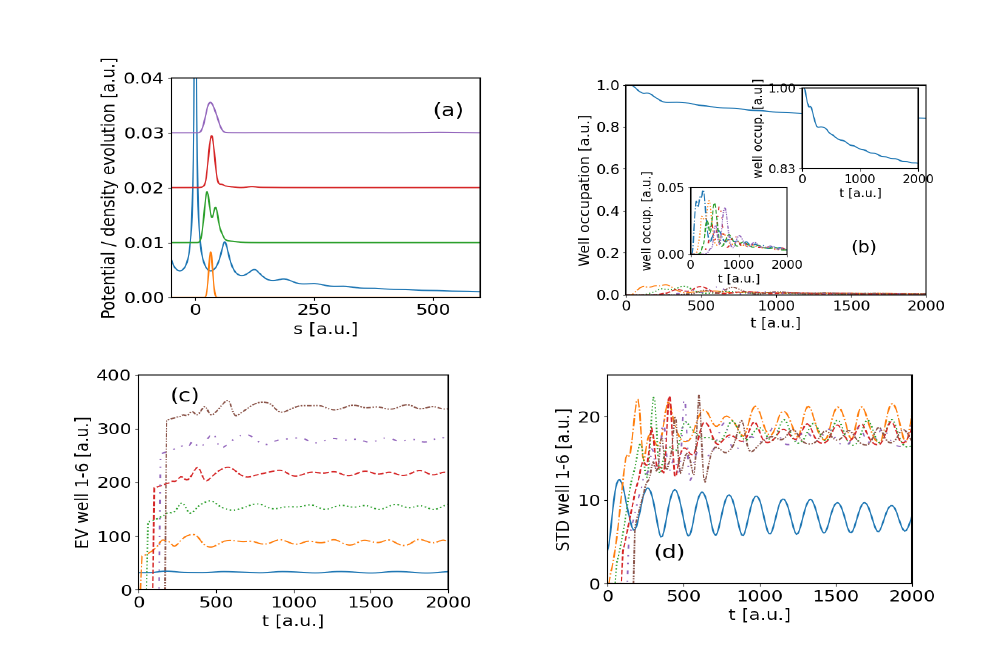}
\caption{(a) The helical potential (h=10,R=10) possessing six distinct minima and snapshots of the density time evolution
for times $t=0,100,200,1000$ (from bottom to top) for an initial Gaussian wave packet located at $s_0 = 32.5$
with zero momentum and width $ \Delta s = 4$ for $M=1$. The underlying sine DVR grid has 3301 points in the interval
$[-150,1500]$. Note that both the potential and the densities are scaled (and shifted)
to fit the same coordinate scale. (b) The evolution of the density integrated separately and individually
over the first innermost to the sixth outermost well of the potential.
The linestyles for the first to the sixth well probabilities are solid, dash dotted, dotted, dashed, dash dot dotted 
and densely dash dot dotted. The insets zoom into relevant regimes of evolution to identify the corresponding
behaviour. (c,d) show observables of the intrawell dynamics: the expectation values of the coordinate $s$ and
its standard deviation, respectively, as a function of time for the six individual wells. Line style as in (b).}
\label{Fig:9}
\end{figure*}

\noindent
Next our setup is the same initial WP like above, but we position it at $s_0 = 32.5$
which is the center of the first deep potential well. The DWP evolution is shown in
Fig.\ref{Fig:9}(a) for $t= 0,100,200,1000$. There is one main broad peak which persists
for all time, i.e. the main part of the WP is trapped in the innermost well; however,
it has a nontrivial dynamics and develops in a transient manner a double peak structure.
The corresponding IWO in Fig.\ref{Fig:9}(b) shows that the IWO for the innermost well
decreases monotonically over the considered time interval from $1.0$ to approximately $0.83$. 

\noindent
Via the inset it can be observed that this decay exhibits some substructure i.e. it shows
a sequence of miniplateaus that correspond to a 'pulsed emission behaviour'
of the intrawell dynamics of the first well. The second inset shows the behaviour of the higher order IWOs
which show a steep short time raise in consecutive order followed by a tail. The latter reflects
the miniplateau behaviour upon closer inspection.

\noindent
Fig.\ref{Fig:9}(c,d) address the intrawell dynamics by inspecting the (renormalized) expectation 
value $<s>$, and the corresponding standard deviation $\sqrt{<s^2>-<s>^2}$, respectively, as a function
of time for all six wells.
The evolution of the expectation values indicate that there is a center of mass
motion taking place with varying amplitude from well to well. The innermost well center of mass dynamics possesses
the smallest amplitude. The different oscillations are not synchronized and exhibit a slightly irregular character.
For the standard deviation of the innermost well we observe a very regular almost single frequency
oscillation around a mean value of eight indicating that a strong breathing dynamics takes place.
For the second to sixth well
the standard deviation dynamics shows an irregular short time behaviour followed an approximately
single frequency regular breathing dynamics around a mean value of eighteen for longer times.

\begin{figure}[H]
\centering
\includegraphics[width=8cm,height=5cm]{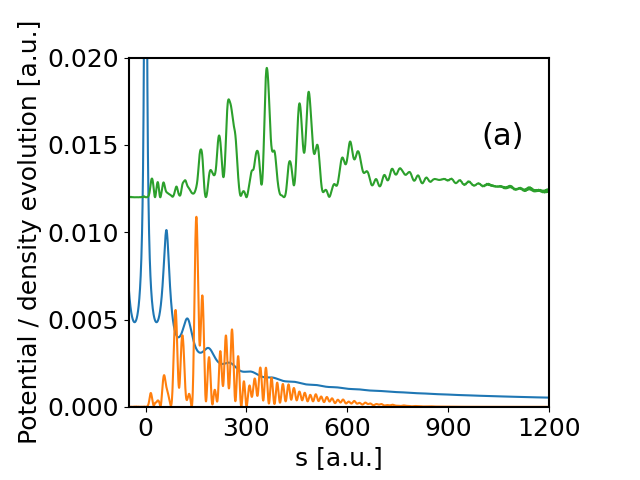}
\includegraphics[width=8cm,height=5cm]{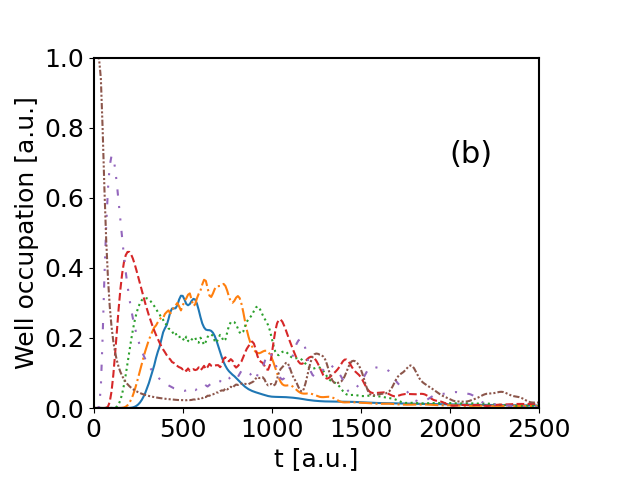}
\caption{(a) The helical potential (h=10,R=10) possessing six distinct minima and snapshots of the density time evolution
for times $t=940,1500$ (from bottom to top) for an initial Gaussian wave packet located at $s_0 = 350.0$
with momentum $p_0=-0.3$ and width $ \Delta s = 4$ for $M=1$. The underlying sine DVR grid has 3301 points in the interval
$[-150,1500]$. Note that both the potential and the densities are scaled (and shifted)
to fit the same coordinate scale. (b) The evolution of the density integrated separately and individually
over the first innermost to the sixth outermost well of the potential.
The linestyles for the first to the sixth well probabilities are solid, dash dotted, dotted, dashed, dash dot dotted 
and densely dash dot dotted.}
\label{Fig:10}
\end{figure}

\noindent
Fig.\ref{Fig:10}(a) shows the DWP evolution for a case with nonzero negative initial momentum
$p_0=-0.3$ for the case $s_0 = 350.0$ at times $t=940,1500$. Again we observe a rich transient
structure formation in the pulse: at $t=940$ we have for short distances a series of double
peaks whose amplitudes increase with increasing distance $s$ followed by several beats on top
of a nonzero background. At later times $t=1500$ this pulse has changed its appearance and
shows now several stretched much broader humps with multiple sharp peaks on top followed by 
a background which exhibits small amplitude undulations. 
The IWOs in Fig.\ref{Fig:10}(b)
show now a much more complex oscillatory behaviour reflecting the DWPs structure formation:
upon a first rise in the course of the WP propagation multiple suboscillations can be encountered
changing the detailed profile of the decreasing envelope for all IWOs.

\subsection{Quantum dynamics in the helical landscape for $h=10,R=10,M=10$}
\label{qdcase3}

\noindent
So far we have been focusing on the case $M=1$, but the mass is an important parameter for
determining the number of bound states in the individual wells. To increase the latter we consider
now the situation $M=10$ for the helix with $h=10,R=10$, i.e. for six wells.
Then the first to sixth (isolated) well possess 14,7,2,2,2,0 bound states, respectively.
We therefore expect a rich time evolution with a pattern forming interference-based wave packet dynamics.

\begin{figure}[H]
\centering
\includegraphics[width=8cm,height=6cm]{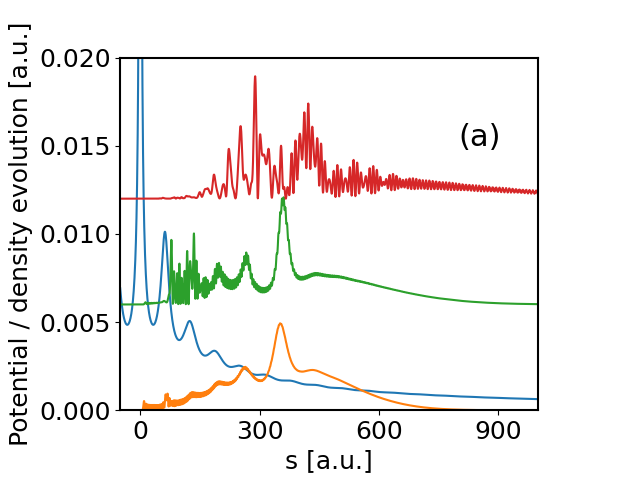}
\includegraphics[width=8cm,height=6cm]{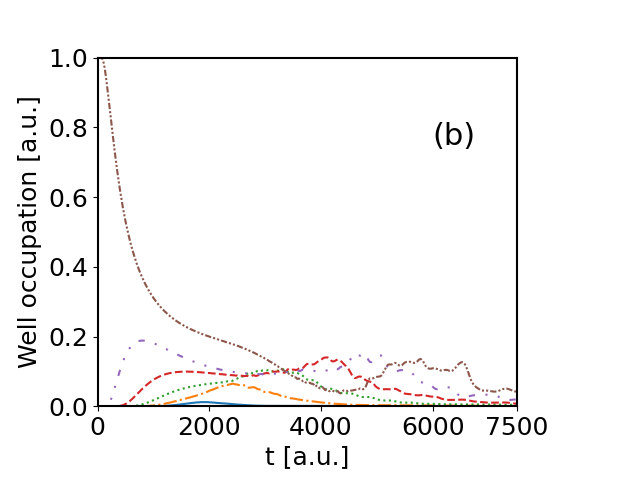}
\caption{(a) The helical potential (h=10,R=10) possessing six distinct minima and snapshots of the density time evolution
for times $t=1800,2500,5000$ (from bottom to top) for an initial Gaussian wave packet located at $s_0 = 350.0$
with zero momentum and width $ \Delta s = 1.2$ for $M=10$. The underlying sine DVR grid has 3301 points in the interval
$[-150,1500]$. Note that both the potential and the densities are scaled (and shifted)
to fit the same coordinate scale. (b) The evolution of the density integrated separately and individually
over the first innermost to the sixth outermost well of the potential.
The linestyles for the first to the sixth well probabilities are solid, dash dotted, dotted, dashed, dash dot dotted 
and densely dash dot dotted.}
\label{Fig:11}
\end{figure}

\noindent
Fig.\ref{Fig:11}(a) shows the DWP evolution for the times $t = 1800, 2500, 5000$ for an initial
WP located at $s_0 = 350$ with zero momentum and width $\Delta s = 1.2$. At $t=1800$ there is, analogously
to the above case for $M=1$ and the same helical landscape a dominant central broad peak. Towards larger
distances we have a smooth decay but for shorter distances the well-structure of the potential
leaves its fingerprints in the oscillations of the DWP.

\noindent
Additionally there is now a dense very high
frequency modulation superimposed on the DWP with moderate amplitude. At later times ($t=2500$) the
innermost part of the DWP starts restructuring on top of an increased amplitude modulation. Step by step
in the time evolution this happens for all oscillatory structures to the left of the main peak
until we finally observe a complete reshaping of the pulse. 
From shorter to larger distances the pulse
has now a sequence of low frequency modulated broak peaks with increasing amplitude followed by a
very broad main peak all with high frequency superimposed undulations and a sequence of beats of low
amplitude forming the tail of the DWP. Again, this structure moves to larger distances as a whole
while roughly maintaining its shape for some transient intermediate times (not shown here). 
Fig.\ref{Fig:11}(b) shows the evolution of the corresponding IWO being in all major aspects similar to the above-observed
one for $M=1$ with individual time evolutions, of course, differing.

\begin{figure*}[htpb!]
\centering
\includegraphics[width=\textwidth]{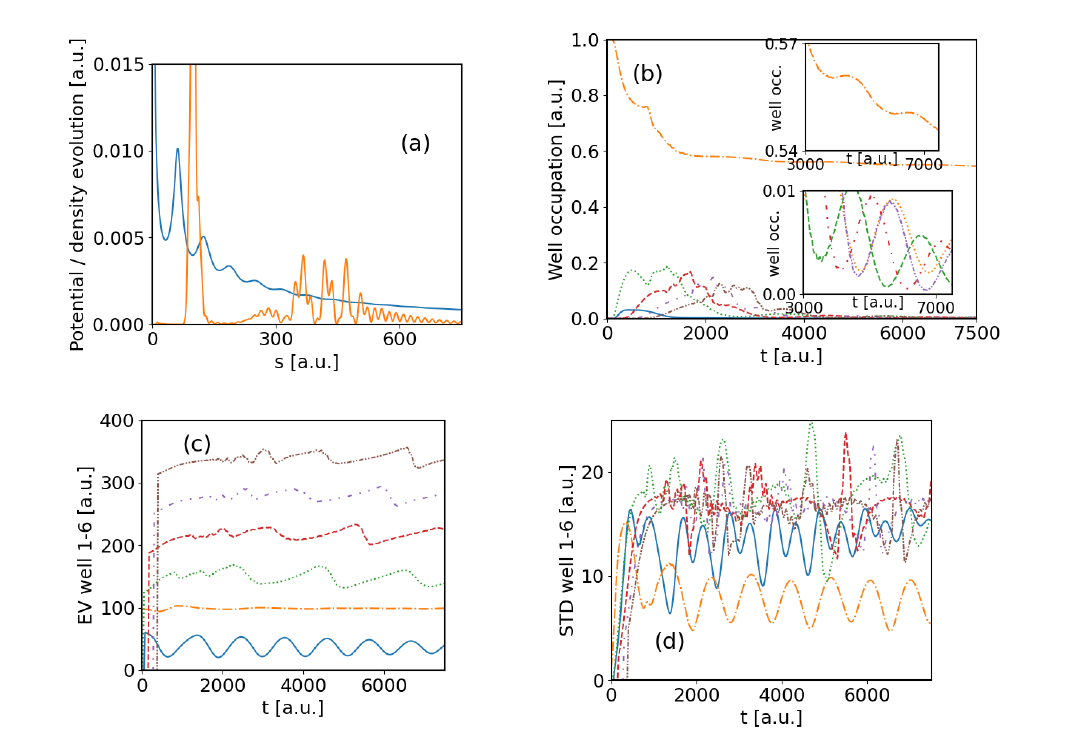}
\caption{(a) The helical potential (h=10,R=10) possessing six distinct minima and a snapshot of the density time evolution
for time $t=3000$ for an initial Gaussian wave packet located at $s_0 = 98.0$
with zero momentum and width $ \Delta s = 1.2$ for $M=10$. The underlying sine DVR grid has 3301 points in the interval
$[-150,1500]$. Note that both the potential and the densities are scaled (and shifted)
to fit the same coordinate scale. (b) The evolution of the density integrated separately and individually
over the first innermost to the sixth outermost well of the potential.
The linestyles for the first to the sixth well probabilities are solid, dash dotted, dotted, dashed, dash dot dotted 
and densely dash dot dotted. The two insets shows zooms into corresponding regions to identify the behaviour.
(c,d) show observables of the intrawell dynamics: the expectation values of the coordinate $s$ and
its standard deviation, respectively, as a function of time for the six individual wells. Line style as in (b).}
\label{Fig:12}
\end{figure*}

\noindent
It is now, for $M=10$, instructive to explore the case for which the initial WP is centered in a
potential well of the helical landscape, specifically the second well.
To be concise, Fig.\ref{Fig:12}(a) shows a single snapshot at
$t=3000$ for the parameter values $s_0 = 98.0$ and $\Delta s = 1.2$. What we can observe in this DWP
snapshot is a broad peak with a thin shoulder towards larger distances which indicates the dynamics
taking place in this second well of the major fraction of the WP remaining there within the finite
time evolution considered here. Also clearly visible is the prominent pulse that has been emitted
from the initial WP up to this time $t=3000$ and is mainly localized in the domain $200<s<700$.
It shows three central main peaks and both, left and right to them, a decaying profile with small
amplitude oscillations on top.

\noindent
Since this is only a snapshot further intriguing dynamics happens
in the course of the time evolution, as can be seen from the IWO dynamics in Fig.\ref{Fig:12}(b).
First of all the IWO for the first innermost well is overall small and only essentially nonzero
for short times, which is a consequence of the still well-separated character of the innermost
well from the outer ones for the given WP energy. 

\noindent
The dominant IWO for the second well rapidly
decays from the initial value very close to one, and this decay slows temporarily down around
$t \approx 800$ before it accelerates again and reaches out to an asymptote for $t > 2000$.
But even during this long-time behaviour (see inset in Fig.\ref{Fig:12}(b)) the second IWO 
shows a sequence of plateaus in its decay which stem from the pulsed emission of probability
from the second well due to its internal dynamics (see below). The IWO for the third and further
outer wells up to the sixth well show a strong participation in the dynamics for the time interval
$0<t<4000$. However, even beyond this time interval the corresponding inset in Fig.\ref{Fig:12}(b)
demonstrates its interesting behaviour: it exhibits oscillations with a certain phase relation
related to their spatial separation. These oscillations are, of course, directly related to the
previously mentioned pulsed emission from the IWO dynamics of the second well.

\begin{figure}[H]
\centering
\includegraphics[width=8cm,height=6cm]{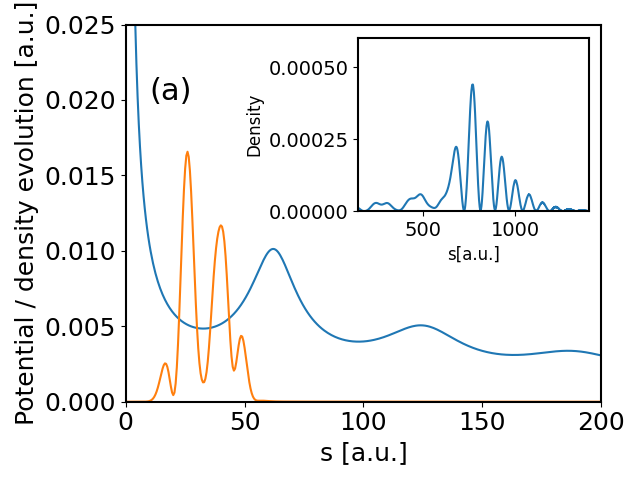}
\includegraphics[width=8cm,height=6cm]{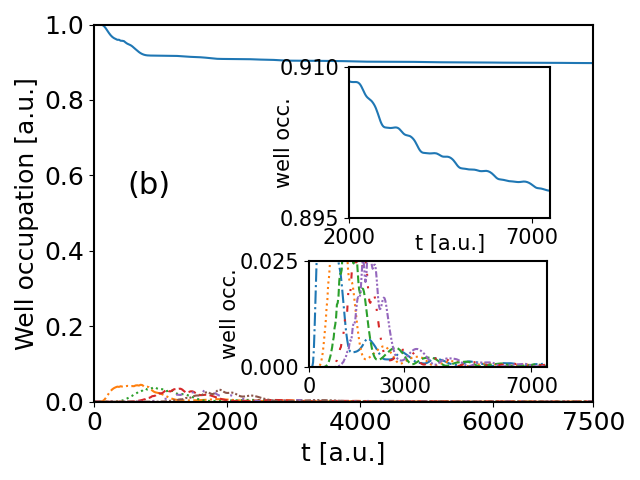}
\caption{(a) The helical potential (h=10,R=10) possessing six distinct minima and a snapshot of the density time evolution
for time $t=4100$ for an initial Gaussian wave packet located at $s_0 = 32.5$
with zero momentum and width $ \Delta s = 1.2$ for $M=10$. The underlying sine DVR grid has 4001 points in the interval
$[-500,1500]$. Note that both the potential and the densities are scaled (and shifted)
to fit the same coordinate scale. The inset shows a zoom into the tail of the wave packet
to resolve its structure. (b) The evolution of the density integrated separately and individually
over the first innermost to the sixth outermost well of the potential.
The linestyles for the first to the sixth well probabilities are solid, dash dotted, dotted, dashed, dash dot dotted 
and densely dash dot dotted. The insets zoom into relevant regimes to identify the underlying behaviour.}
\label{Fig:13}
\end{figure}

\noindent
Fig.\ref{Fig:12}(c,d) addresses the intrawell dynamics i.e. the expectation values of the coordinate $s$ and
its standard deviation, respectively, as a function of time for the six individual wells. 
The expectation value of the innermost well shows a single frequency harmonic oscillation,
the second well shows almost no dynamics at all at least for not too short times,
and the higher order wells show similar very anharmonic oscillations with a strong asymmetry
w.r.t. their forward (increasing $<s>$) and backward (decreasing $<s>$) motion.
The standard deviation shows a few mode dynamics for the first and second well and 
a much more irregular multi-mode dynamics for the higher wells. On top of small amplitude
oscillations there occur narrow peaks of intermittent dynamical character. 

\noindent
Finally, let us inspect the DWP evolution for a WP initially centered in the first innermost well
with $s_0 = 32.5$ and $\Delta s = 1.2$. A snapshot for $t = 4100$ is shown in Fig.\ref{Fig:13}(a).
The density profile in this first well now shows at this time multiple large amplitude
oscillations on top of a broad distribution. This, of course, heavily varies with changing time
due to the strong internal well dynamics (see below). Again, the inset shows a pulse that has
been emitted previously with low amplitude overall and characteristic shape, which is traveling
with increasing time to larger distances. Other pulses do follow regularly, as can be seen from
the corresponding IWOs in Fig.\ref{Fig:13}(b), see in particular the inset for the IWO evolution of the
first well showing a characteristic plateau structure. The IWOs of the second to higher wells are,
compared to the case of Fig.\ref{Fig:12} somewhat suppressed but they exhibit a similar characteristic
behaviour, as seen in the corresponding inset of Fig.\ref{Fig:13}(b).

\section{Conclusions and Outlook}
\label{concl}

\noindent
The idea of exploring long-range interacting particle systems on curved manifolds for which the
particles are dynamically constrained to move on a lower-dimensional surface but interact via the
entire three-dimensional space has by now proven to lead to a variety of interesting structures
and phenomena, as well as an intriguing dynamics \cite{Schmelcher11,Doerre25,Zampetaki15-1,Zampetaki15-2,Zampetaki17,Zampetaki18,
Plettenberg17,Siemens20,Siemens21,Gloy22,Pedersen14,Pedersen16-1,Pedersen16-2,Stockhofe14,Zampetaki13}.
However, the absolute majority of these works are for classical particle systems and we present
here the first exploration of the quantum two-body problem for two charged particles on a helix.
The helix is of particular interest for several reasons: due to the constant curvature and 
torsion (i) it allows for a simple quantum Hamiltonian structure (ii) it leads to
a separation of the center of mass and internal motion (iii) the tunable number of (classically stable)
equilibria and wells leads to a rich structural, and consequently dynamical, characteristics
between symmetry-dominated order and complete disorder. For the repulsive two-body problem this 
implies that the equilibria of the two particles are (approximately)
on opposite sides of the same or different windings. For more particles this picture is 
systematically modified and enhanced and leads to an intriguing cluster formation \cite{Doerre25}.

\noindent
Here we perform the first steps towards an understanding of the quantum properties of repulsively
interacting helical particle systems at hand of the two-body case which reduces, after separation
of the center of mass, to a pure relative motion problem. It turns out that, additionally to the
well-known fact that the number of potential wells can be changed by tuning the pitch $h$ and radius $R$,
the anharmonicity of the individual wells varies strongly. The relevant parameter to change the degree
of anharmonicity is $\frac{h}{R}$ and, besides this, the order of the well in a given helical potential.
Correspondingly we inspected the individual (isolated) wells spectra and found that the energy eigenvalue spacing
can be approximately constant, i.e. harmonic oscillator-like, over a broad range of excitations to strongly 
increasing due to a strong anharmonicity.

\noindent
The main theme of this work is the study of the quantum dynamics i.e. time evolution of wave packets
with varying location in the multi-well potential landscape, varying width and momentum. We have hereby
been focusing on two setups with three versus six potential wells. We also considered both cases
$M=1$ and $M=10$. For $M=1$ the setup with $h=5.8,R=4.0$ (three wells) possesses a single bound
state for the isolated inner well and none for the outer ones. This case was, in a nutshell, the
one with the, compared to the other cases, least complex quantum dynamics. Nevertheless, we observed
a restructuring of the wave packet and the formation of beats or simply a fragmentation of the
wave packet upon initially localizing it between two wells. The tunneling out of a well appeared
to be rather structureless, which is not unexpected due to the fact that the well had one or no
bound states.

\noindent
For the helical potential $h=10,R=10$ and $M=1$ with six wells with a series of bound states and many
for $M=10$ the interference dynamics becomes significantly richer in
pattern forming pulses. For the initially far off the series of wells located WP the latter are
imprinted on the WP while spreading towards the repulsive Coulomb branch and exhibit additional
short wavelength smaller amplitude oscillations. A major restructuring of this pulse occurs while
completing the reflection dynamics into a sequence of broader peaks followed by a sequence of beats
with decreasing amplitude, which then, transiently, travels as a complete pulse to larger distances.
Locating the initial WP at an inner well leads to a rich above barrier and tunneling dynamics due to the intrawell
dynamics: pulse shaping is combined with an overall oscillating emission of them. We have analyzed
these features, among others, by inspecting the individual well occupation and the intrawell dynamics.

\noindent
Possible experimental preparations of helices are state of the art nanostructure fabrication
techniques based on self-organization or lithography. In this case the electron or hole dynamics
would be on the scale of tens/hundreds of nanometers. Ultracold trapped ions are a second promising
candidate. Here it would need a design of helical field configurations on the micrometer scale to trap the previously laser cooled
atomic ions thereby reaching sufficiently low temperatures to enter the quantum regime. 

\noindent
While the present work shows the variety of possibilities for the pattern and pulse shape forming reflection
dynamics in the collision of the two charged particles while propagating across the windings, it is only the
beginning of the exploration of the helical quantum problem. An immediate next step would be the
investigation of the more particle dynamics which carries the promise of showing a variety of 
transient binding schemes and would involve many different pattern formation time scales.
Alternatively and complementary to the present quantum dynamical approach, and in order
to compute the underlying resonances, i.e. their energetical positions and widths, a complex
scaling approach might be developed and applied.

\section{Acknowledgments}
\label{ack}

The author acknowledges many helpful discussions with H.D. Meyer concerning the
usage of the Heidelberg MCTDH software package. This work has been in part conducted during a 
visit to the center for Theoretical Physics at Hainan University whose hospitality
is greatly appreciated.

\end{document}